\documentclass{iaus}
\usepackage{graphicx}

\title[Complexity in small-scale dwarf spheroidal galaxies] %% give here short title %%
{Complexity in small-scale dwarf spheroidal galaxies}

\author[A. Koch et al.]   %% give here short author list %%
{Andreas Koch$^1$, Daniel Ad\'en$^2$,  Eva K. Grebel$^3$, \and Sofia Feltzing$^2$}

\affiliation{$^1$University of Leicester, University Road, LE1 7RH Leicester, UK \\ email: {\tt ak326@astro.le.ac.uk} 
 \\[\affilskip] $^2$Lund Observatory, Box 515, SE-75120 Uppsala, Sweden % \\ email: {\tt daniela,sofia@astro.lu.se}
 \\[\affilskip] $^3$Astronomisches Rechen-Institut, M\"onchhofstrasse 12-14, 69120 Heidelberg, 
 Germany }%\\ email: {\tt daniela,sofia@astro.lu.se}

\pubyear{2009}
\volume{265}  %% insert here IAU Symposium No.
\pagerange{119--126}
\jname{Chemical Abundances in the Universe: Connecting First Stars to Planets}
\editors{K. Cunha, M. Spite \& B. Barbuy, eds.}
\begin{document}

\maketitle

\begin{abstract}
Our knowledge about the chemical evolution of 
the more luminous dwarf spheroidal (dSph) galaxies is constantly growing. 
However,  little is known about the enrichment of the ultrafaint systems 
recently discovered in large numbers in large Sky Surveys.  
Low-resolution spectroscopy and photometric data indicate that these 
galaxies are predominantly metal-poor. On the other hand, the most recent 
high-resolution abundance analyses indicate that some of these galaxies 
experienced highly inhomogenous chemical enrichment, where star formation 
proceeds locally on the smallest scales. Furthermore, these galaxy-contenders 
appear to contain very metal-poor stars with [Fe/H]$<-3$ dex and could 
be the sites of the first stars. 
Here, we consider the presently available chemical abundance information 
of the (ultra-) faint Milky Way satellite dSphs. In this context, some of the most 
peculiar element and inhomogeneous enrichment patterns will be discussed and related
to the question of to what extent the faintest dSph candidates  
and outer halo globular clusters  could have contributed to the metal-poor Galactic halo. 
\end{abstract}
\firstsection % if your document starts with a section,%
\section{Introduction}
Dwarf spheroidal (dSph) galaxies are intriguing for a plentitude of reasons: 
Owing to their very low luminosities ($M_V$$\ge$$-14$ mag) they have been characterized as faint systems ever since their first discovery. 
They further have low total masses of only a few 10$^7$ $M_{\odot}$  and a puzzling deficiency of gas (e.g., Grebel et al. 2003; Bailin \& Ford 2007; Gilmore et al. 2007). 
Over the past three years, the faint end of the galaxy luminosity function has been traced even  further down, towards the {\em ultrafaint} regime. 
These ``ultrafaint'' dSphs, discovered in large number in sky surveys such as the SDSS, are now the faintest galaxies known to exist in the Universe, 
with absolute magnitudes above $M_V > -6$,  and stellar masses of up to a mere few ten thousand Solar masses 
(e.g., Martin et al. 2008).  

Furthermore, the dSphs are fairly  metal-poor systems, with mean metallicities starting at $-$1 to $-$2 dex and decreasing. 
The range of metallicity in a given dSph is normally large and of up to 0.5 dex. 
Typically these spreads greatly exceed the measurement errors.  
Despite deep photometric studies and complementing spectroscopy for selected stars, the detailed properties of 
these ultrafaint galaxies remain poorly investigated until now. 
Interestingly, the ultrafaint dSphs are more metal-poor on average than 
their   more luminous counterparts; their mean metallicities reach as low as about  $-2.5$ dex (e.g., Simon \& Geha 2007).
While no star more metal-poor than [Fe/H]$< -3$ dex had been found in any of the classical dSphs until recently (e.g., Koch et al. 2006; Helmi 2006; 
cf. Cohen \& Huang 2009), several such metal-poor stars,  down to $-3.3$ dex  have been detected in the ultrafaint galaxies (Kirby et al. 2008; Norris et al. 2008; 
Frebel et al. 2009).  

Although cosmological simulations like $\Lambda$CDM predict a wealth of small-scale substructures that hierarchically merge 
into larger structures like the present-day Milky Way (MW), 
a number of arguments against such a simplistic 
view has arisen over the years (Moore et al. 1999). 
Those comprise the oft-cited 
missing satellite problem, which is, however, nowadays much alleviated 
(e.g., Robertson et a. 2005; 
Simon \& Geha 2007).  
Another problem is the discrepancy between the chemical abundances of the dSph stars compared to the halo stars (Sect.~2). 
Furthermore, the aforementioned apparent lack of very metal-poor stars in the dSphs was long considered 
a major contradiction to the large number of such stars, below $-$3 dex, found in the Galactic halo.  
This leaves us with the question of how and when the 
(ultrafaint) dSphs formed and evolved,  and how they fit 
into the cosmological $\Lambda$CDM models. 
In particular, what fraction of dSph-like systems contributed to the build-up of the stellar halo of the MW? 
\vspace{-3ex}
\section{Chemical abundances -- The general picture}
In Fig.~1 we show the [Ca/Fe] ratio, as an example of the $\alpha$-element distribution, for the currently available data for dSphs  (see Koch 2009 for a detailed review 
and the source of those data), in comparison to the Galactic disks and local halo stars. 
\begin{figure}[htb]
% \vspace*{-2.0 cm}
\begin{center}
 \includegraphics[width=0.7\hsize]{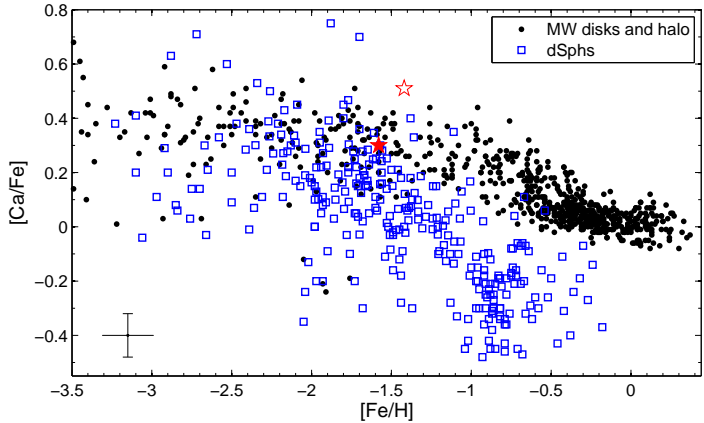} 
% \vspace*{-1.0 cm}
 \caption{Currently available abundance data for Galactic stars (black dots) and dSphs (squares). See Koch (2009) for a detailed review of the sources for these data. 
 Red symbols indicate the outer halo GCs Pal~3 (Koch et al. 2009) and Pal 4 (Koch \& C\^ot\'e in prep.)}
   \label{fig1}
\end{center}
\end{figure}
Since the first observations by Shetrone et al. (2001) that the dSph stars are systematically depleted 
in those elements relative to halo stars at the same metallicity  (as already predicted by the models of Unavane et al. 1996), the abundance data 
have now vastly grown, with up to several tens of stars in a few of the 
more luminous systems (e.g., Koch et al. 2008a, 2009; Tolstoy et al. 2009) and progressive measurements in 
the fainter ones (e.g., Koch et al. 2008b; Frebel et al. 2009). The first thing to note is that 
this picture of $\alpha$-depletion remains valid also when taking the new data into account. 
However, the new picture  that slowly emerges is that there are in fact metal-poor stars below $-3$ dex found 
in the classical dSphs (Cohen \& Huang 2009;  Frebel et al., this meeting [S265-o:18]) and in particular the ultrafaint dSphs appear to host a large number of those 
stars (Kirby et al. 2008; Norris et al. 2008; Frebel et al. 2009). For many of those stars, high-resolution abundance data are currently being gathered.  
The overlap of these metal-poor stars' abundances with those of the Galactic metal-poor halo plateau at [$\alpha$/Fe]$\sim$+0.4 dex 
then underlines the picture in which the accretion and disruption of dSph-like systems had a major contribution to the {\em metal-poor halo at most}.  
\vspace{-3ex}
\section{A case study -- the Hercules dSph}
Hercules (hereafter Her) is one  of the ``ultrafaint'' dSph galaxies discovered in the SDSS (Belokurov et al. 2007). 
Past studies have established a low mean metallicity and indications of a low mass and a high mass-to-light ratio (Simon \& Geha 2007).
\subsection{Chemical element abundances}
In Koch et al. (2008b) we showed that the elemental abundance patterns in Her are peculiar. Firstly, neither of the only two red giants analyzed in the 
literature to date show any evidence for heavy elements (e.g., Ba, Eu) above the noise in the spectra.
 Secondly, we found remarkably high abundance ratios of the hydrostatic 
(O, Mg) to the explosive  $\alpha$-elements (Si, Ca, Ti), see Fig.~2. This suggests that  very massive stars were the main drivers for the chemical 
enrichment in Her. For instance, the high [Mg/Ca] of 0.6--1.0 dex in the Her stars suggests that stars of at least 30 M$_{\odot}$ governed the 
enrichment (Heger \& Woosley 2008). Thus we may be seeing the results of stochastical enrichment in terms of an incompletely sampled IMF. 
Such abnormally high [Mg/Ca] ratios are also found in the low-mass dSph Boo~I (Feltzing et al. 2009). 
\begin{figure}[htb]
% \vspace*{-2.0 cm}
\begin{center}
 \includegraphics[width=0.49\hsize]{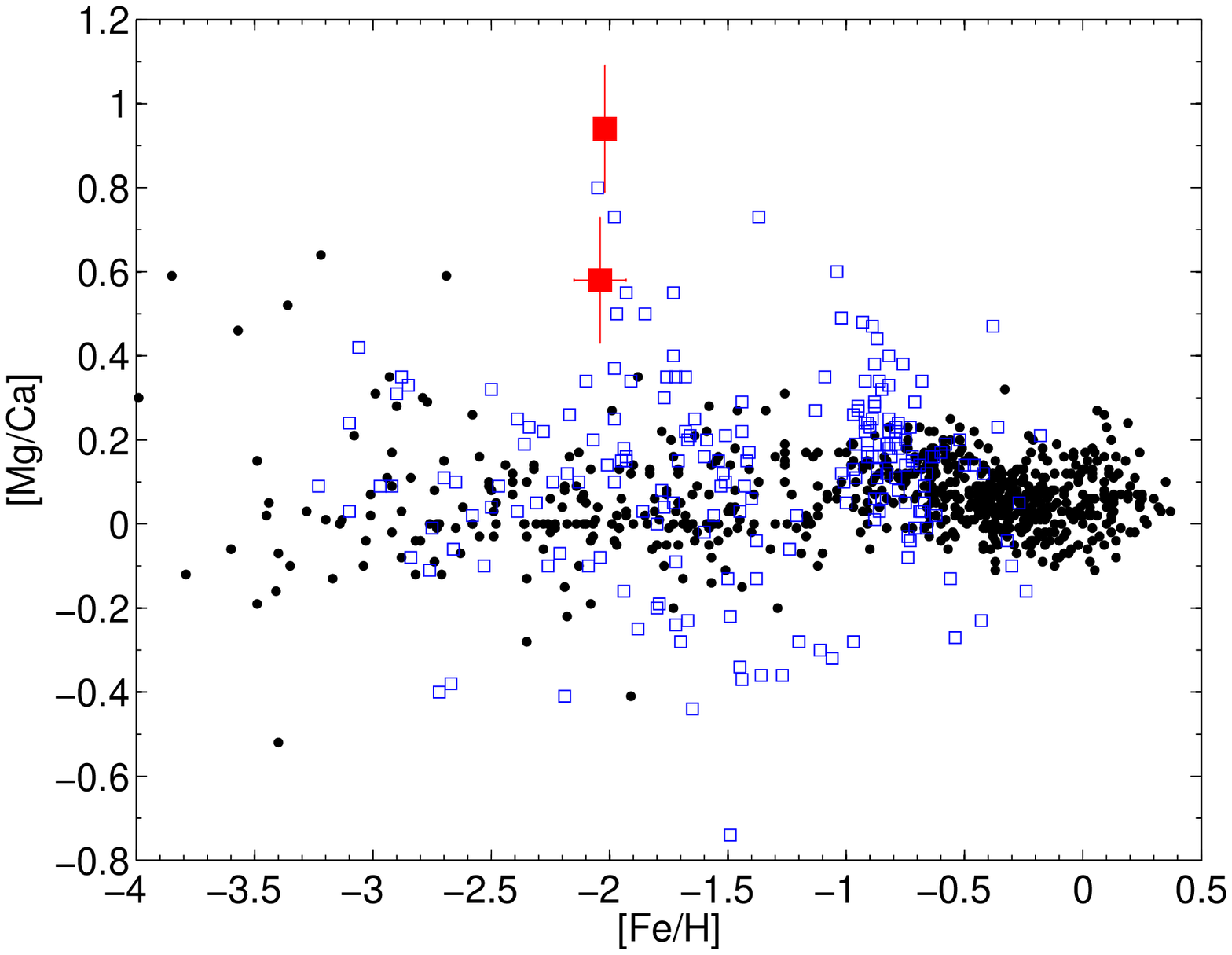} 
  \includegraphics[width=0.49\hsize]{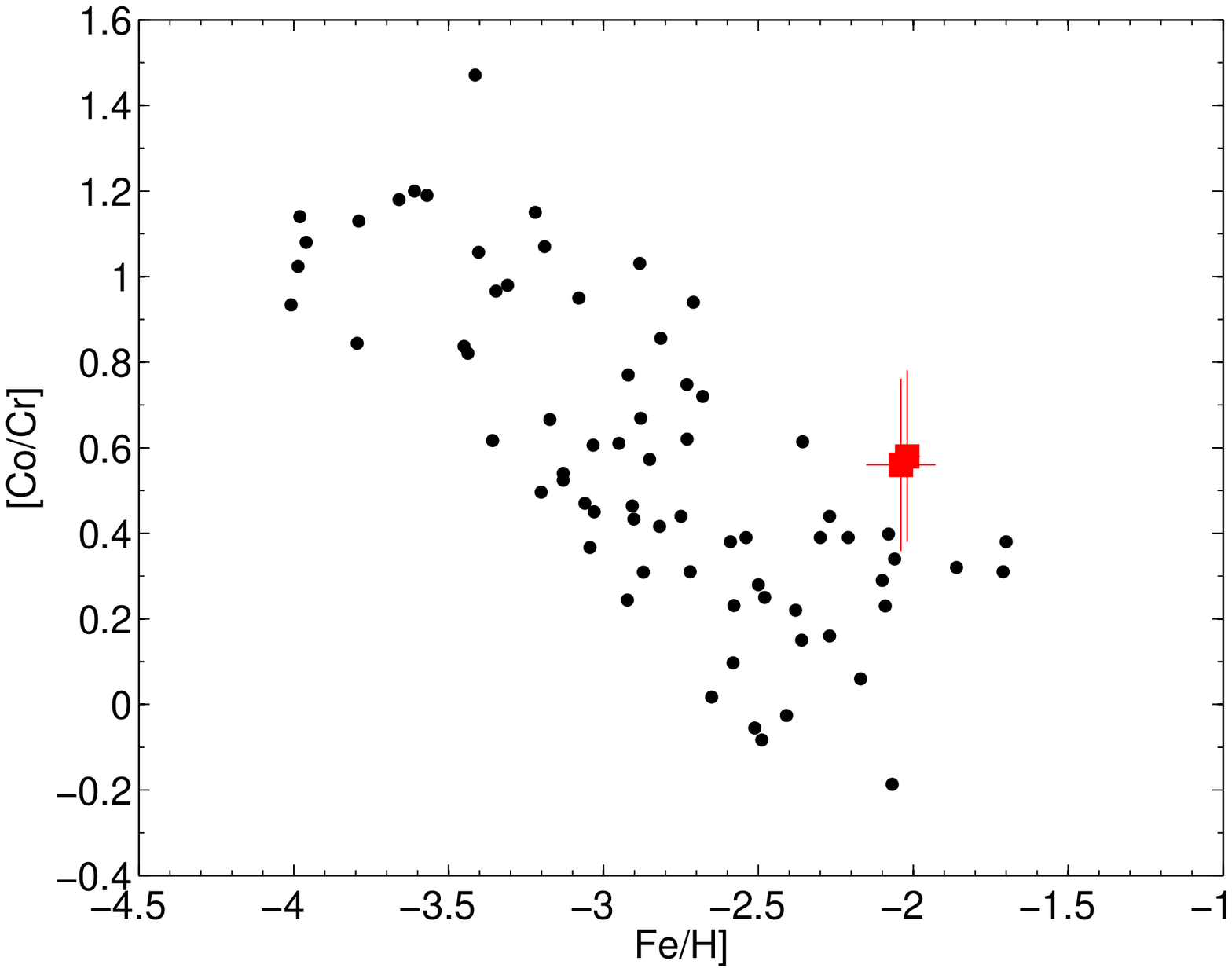} 
% \vspace*{-1.0 cm}
 \caption{Left panel: [Mg/Ca] ratios for the same MW and dSph stars as in Fig.~1. Highlighted as red squares are the two stars in Her. The right panel shows the 
 Her [Co/Cr] abundance ratio in comparison to the metal-poor Galactic halo stars of McWilliam et al. (1995).}
   \label{fig1}
\end{center}
\end{figure}

Another peculiarity is the unusally high 
[Co/Cr] ratio (Fig.~2): at [Co/Cr]$\sim$0.58 dex, the two moderately metal-poor Her stars ([Fe/H] of $-$2 dex) rather resemble the metal-poor Galactic halo below 
$\sim$$-2.5$ dex. This is explicable, if we assume that Her was enriched towards its observed higher Fe-abundances through standard SNe~Ia contributions according 
to its IMF, but it experienced an early enrichment from a first generation of metal-free, very massive Population III stars, the models of which reproduce  
the high [Co/Fe] and low  [Cr/Fe] ratios very well.  
If this scenario is supported by further observations in other (fainter) stars in Her and other dSphs, it would mean that the ultrafaint dwarfs may in fact 
be the site of the very first generations of stars in the Universe.
\subsection{Clean member selection}
One difficulty in the interpretation of resolved galaxy properties from low- or medium resolution data is the inevitable contamination with Galactic foreground stars. 
While the member selection in some dSphs is straightforward, thanks to their high radial velocities of (positive or negative) several hundred km\,s$^{-1}$ 
(e.g., Draco, Carina) 
and/or high Galactic latitudes, pure color-magnitude and radial velocity criteria in low-resolution mode fail for those cases 
where the dSphs are deeply embedded in the foreground. 
With a systemic mean velocity of 45 km s$^{-1}$, Hercules is in fact strongly affected. 

In Ad\'en et al. (2009a) we showed, however, that the dwarf contamination can be efficiently identified using Str\"omgren photometry: this filter system is 
 able to discern the evolutionary stages of stars, based on a set of gravity sensitive index definitions (e.g., Faria et al. 2007). 
 Using this we removed all the contaminating foreground dwarf stars. As a result, we could 
isolate a bona fide  member sample of 45 red giants, AGB, RHB, and BHB stars. As it turned out, about five of the stars that overlap with previous studies 
(Simon \& Geha 2007; Kirby et al. 2008) are likely foreground stars. The 
cleaned member candidate sample then yields a lower velocity dispersion and thus significantly lower mass by a factor of $\sim$3 (Ad\'en et al. 2009b) compared to the literature (e.g., Simon \& Geha 2007; Strigari et al. 2008). 
Furthermore, our sample is slightly more metal-rich than the previous estimates. Both from our calcium triplet spectroscopy and from the calibration of the 
Str\"omgren photometry onto metallicity, we find a mean [Fe/H] of $-$2.35 dex with a 1$\sigma$ scatter of 0.31 dex.  
Currently, there are four objects known that have systemic radial velocities comparable to the Galactic foreground around 0--50 km\,s$^{-1}$, viz.  
CVn~I, Willman~I, Leo~T, and Her. 
Therefore we emphasize the importance of a clear member selection for all future studies, by means of sophisticated photometric and spectroscopic techniques.  
High-resolution follow-up is vital for assessing the true stellar properties and sampling the true, full metallicity (preferably, iron) ranges of the dSphs. 
\vspace{-3ex}
\section{The outer halo GC Palomar 3}
Now that we argued about the role of the dwarf galaxies for building up parts of the Galactic halo, we can move on to the {\em outermost} halo 
and its globular clusters (GCs). It has long been known that there is a distinct dichotomy in the field star populations of the Milky Way (e.g., Hartwick 1987; 
Carollo et al. 2007) and also M31 (Koch et al. 2008c). Secondly, the lack of a metallicity gradient within the outer halo GC system as well as the occurrence of a pronounced 
second parameter problem had prompted the original scenario of the accretion origin of the Galactic halo by Searle \& Zinn (1978). It is thus natural to ask, 
whether the outermost GCs are either potential building blocks of the halos  or whether those systems have been donated to the halo themselves by 
disrupting dSph-like galaxies. 

In order to look for chemical differences or similarities between those components,  we carried out a spectroscopic study of the remote GC Pal~3 
(R$\sim$92 kpc; e.g., Hilker 2006). This cluster is also one of the most extended GCs of the MW system (r$_h\sim$15 pc), and its proper motion, within its uncertainties,  
does not exclude the possibility that it is not bound to the MW at all. Our sample of 4 red giants, observed with the Magellan/MIKE spectrograph, supplemented 
by integrated abundances of  19 stars targeted with Keck/HIRES, however, showed that Pal~3 bears close resemblance to the majority of 
both inner and outer halo GCs (Koch et al. 2009). Its $\alpha$-elements are enhanced to the halo value of $\sim$0.4 dex (Fig.~1), and its iron peak elements are 
compatible with  Solar values. In fact, 80\% of its abundance ratios are identical to those of the archetypical inner halo GC M~13 within the uncertainties; the same 
holds for a comparison with the outer halo cluster NGC~7492 (Cohen \& Mel\'endez 2005a,b). 
On the other hand, Pal~3 does not resemble dSph field stars in any regard: the enhanced $\alpha$ abundances and most other abundance patterns are incompatible 
with, e.g., the $\alpha$-depletions seen in the dSph stars. 
The situation is, however, slightly different for the GC system of the Fornax dSph (Letarte et al. 2006). Its abundance patterns are unlike the surrounding dSph field stars 
and rather appear to be similar to the patterns found in the Galactic halo and its GCs and a possible resemblance with Pal~3 cannot be refuted by the present data. 
However, the observed very small scatter of the Pal~3 stars advocates against the large abundance spreads detected in all of the Local Group dSphs analyzed to date 
and we conclude that Pal~3 (and Pal~4, Koch \& C\^ot\'e, in prep.) did likely not contribute any major fraction % (of stars and/or GCs?] 
to the outermost halo, but rather are part of an underlying, genuine halo population. 

There is yet one peculiarity found in Pal~3 that deserves notion: All the heavy elements in this cluster are fully compatible with a pure $r$-process 
origin without any need to invoke any significant contribution from the $s$-process. Such a behaviour has so far only been observed in very metal-poor field stars 
(e.g., Honda et al. 2007) and in the GC M~15 (Sneden et al. 2000) and its surrounding stellar stream (Roederer 2009, this meeting [S265-p:8]). Contrary to 
the massive progenitors that dominated the enrichment in the low-mass environment of the Her dSph, the n-capture patterns in Pal~3 do not require any 
such enrichment by massive stars, but are explicable by a standard $r$-process in $\sim$8~M$_{\odot}$ SNe~II (e.g., Qian \& Wasserburg 2003). A detailed interpretation of this loss of s-process material 
is beyond the scope of this work. We just remark that the large spatial extent of this loose cluster and the corresponding shallow potential could have 
favored an early loss of the ejecta of a first generation of AGB stars, followed by efficient star formation with a  high SNe II rate leading to 
strong $r$-enrichment through this second generation of cluster stars (cf. D'Antona \& Caloi 2008). While it is clearly unwarranted to discern any trend of 
the [$r$/Fe] and [$r$/$s$] ratios with structural parameters in Fig.~3, it is certainly noteworthy that Pal~3, 
with its large radius, also exhibits the highest [Eu/Fe]  
(likewise [Dy/Fe]), lowest [Ba/Eu] ratios, respectively.  A complete and homogeneous survey of the n-capture elements within the Galactic GC system is clearly necessary. 
\begin{figure}[htb]
 %\vspace*{-1ex}
\begin{center}
 \includegraphics[width=0.49\hsize]{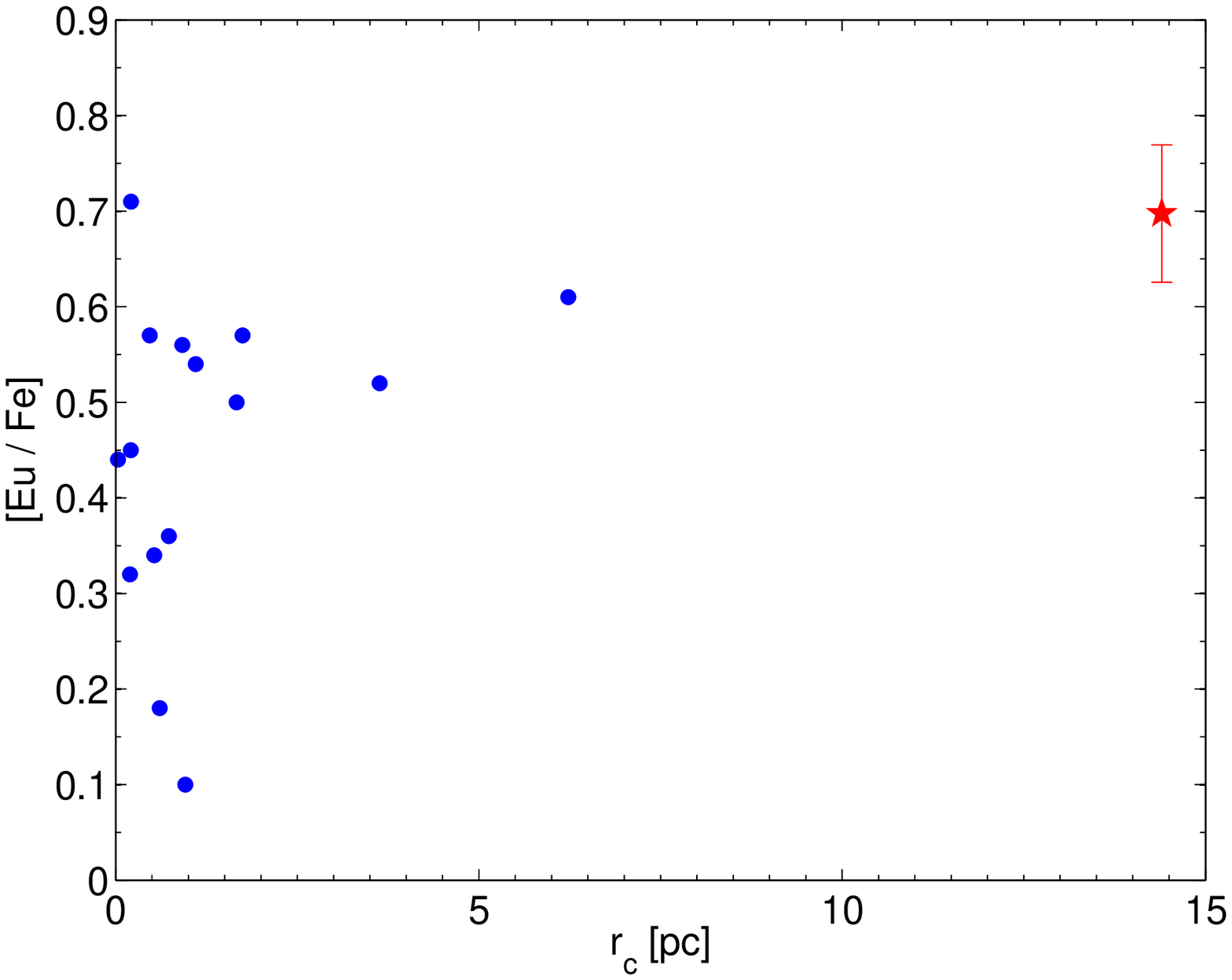} 
 \includegraphics[width=0.49\hsize]{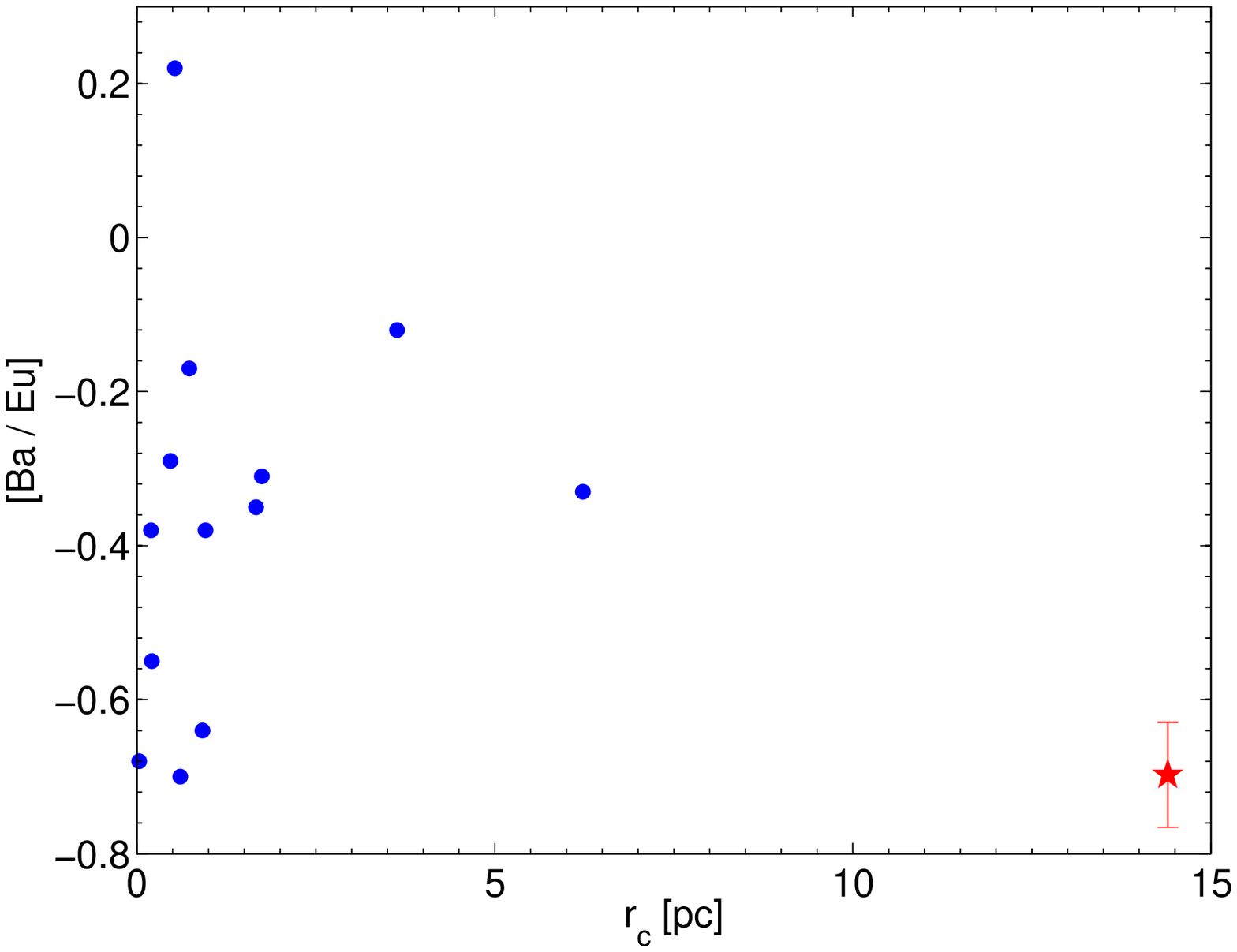} 
% \vspace*{-1.0 cm}
 \caption{Abundance ratios of a sample of Galactic GCs (blue points) as function of their core radii. Pal~3 is denoted as a red star symbol.}
   \label{fig1}
\end{center}
\end{figure}
\begin{discussion}

\discuss{Nomoto}{Your abundance patterns in Hercules could also be explained by enrichment from faint Supernovae with progenitor masses of $\sim$25 M$_{\odot}$ 
[this meeting; S265-i:26].}

\discuss{Koch}{That is an interesting possibility and one should look into fits to the entire abundance distribution of the stars. In any case, 25 M$_{\odot}$ still leaves us \
in the high-mass regime for Her's enrichment.}

\discuss{Sarajedini}{Are you saying from your data in only two GCs that the outer and inner halo are coeval? 
There is evidence that parts of the Galactic GCs were accreted from  dSphs (e.g., the Sagittarius clusters).}

\discuss{Koch}{Our two clusters appear to have experienced a similar chemical evolution as the inner halo clusters. Furthermore, as opposed to the Sgr clusters, 
Pal~3 is not a member of any currently known stream.} 

\end{discussion}

\end{document}